\documentclass[a4paper]{article}

\usepackage{hyperref}

\hypersetup{
    colorlinks=true,
    linkcolor=red,
    filecolor=red,      
    urlcolor=red,
    citecolor=green,
}

\usepackage{INTERSPEECH2021}

\title{Generalized Spatio-Temporal RNN Beamformer for Target Speech Separation}
\name{$^{\dag}$Yong Xu, $^{\ddag}$Zhuohuang Zhang\thanks{This work was done when Z. Zhang was an intern in Tencent.}, $^{\dag}$Meng Yu, $^{\dag}$Shi-Xiong Zhang, $^{\dag}$Dong Yu}
\address{$^{\dag}$Tencent AI Lab, Bellevue, USA ~~~ $^{\ddag}$ Indiana University, Bloomington, USA }
\email{\{lucayongxu,raymondmyu,auszhang\}@tencent.com, zhuozhan@iu.edu}

\begin{document}

\maketitle
\begin{abstract}
Although the conventional mask-based minimum variance distortionless response (MVDR) could reduce the non-linear distortion, the residual noise level of the MVDR separated speech is still high. In this paper, we propose a spatio-temporal recurrent neural network based beamformer (RNN-BF) for target speech separation. This new beamforming framework directly learns the beamforming weights from the estimated speech and noise spatial covariance matrices. Leveraging on the temporal modeling capability of RNNs, the RNN-BF could automatically accumulate the statistics of the speech and noise covariance matrices to learn the frame-level beamforming weights in a recursive way. An RNN-based generalized eigenvalue (RNN-GEV) beamformer and a more generalized RNN beamformer (GRNN-BF) are proposed. We further improve the RNN-GEV and the GRNN-BF by using layer normalization to replace the commonly used mask normalization on the covariance matrices. The proposed GRNN-BF obtains better performance against prior arts in terms of speech quality (PESQ), speech-to-noise ratio (SNR) and word error rate (WER).	
\end{abstract}
\noindent\textbf{Index Terms}: MVDR, Spatio-temporal RNN beamformer, ADL-MVDR, GEV, GRNN-BF, speech separation
\section{Introduction}
\label{sec:intro}

%Purely neural network (NN) based speech separation approaches \cite{wang2014training, luo2019conv, yu2017permutation, xu2014regression}, although can obtain good objective scores, inevitably cause non-linear speech distortions that are harmful for the speech recognition \cite{du2014robust}. On the other hand, the traditional MVDR beamformer with NN-predicted masks \cite{heymann2016neural, erdogan2016improved, heymann2017beamnet, xiao2017time}, although can minimize the non-linear distortions, has limited noise reduction capability \cite{xu2020neural}. 

The mask-based MVDR \cite{heymann2016neural, wang2018mask, erdogan2016improved, heymann2017beamnet, xiao2017time,xu2020neural} could achieve less non-linear distortion than the existing purely ``black box'' neural network (NN) based speech separation methods  \cite{du2014robust, wang2014training, luo2019conv, yu2017permutation, xu2014regression}.
%Complex-valued mask based multi-tap MVDR was proposed in \cite{xu2020neural} to reduce more residual noise. 
However, the residual noise level of mask-based MVDR method is still high \cite{habets2013two, xu2020neural}. %Moreover, due to the numerical instability of covariance matrices, m
Most of mask-based beamformers are optimized in the chunk-level \cite{heymann2016neural, erdogan2016improved, xiao2017time, xu2020neural}. The calculated beamforming weights are hence chunk-level which is not optimal for each frame. Furthermore, the matrix inversion in the traditional beamformer (e.g., MVDR) comes with the numerical instability problem \cite{zhang2021end, chakrabarty2015numerical, lim2017numerical, zhao2012fast}, which is caused by the singularity in the matrix inversion \cite{zhang2021end}. Although this issue could be alleviated by using some techniques, e.g., diagonal loading \cite{mestre2003diagonal, chakrabarty2015numerical, zhang2021end}, it is not fully solved. This problem could be worse in the end-to-end joint training system \cite{zhang2021end, xu2020neural}. Time-varying mask-based beamformers were investigated in \cite{wang2019sequential, kubo2019mask}, however they also have the numerical instability problem \cite{zhang2021end, chakrabarty2015numerical}.

The recurrent neural network (RNN) was once proved to have the ability to solve the matrix inversion \cite{wang1993recurrent, zhang2005design} and the eigenvalue decomposition problems \cite{liu2008recurrent, wang2016recurrent}, which are  two main matrix operations in most of the beamfomers' solutions, e.g., MVDR \cite{erdogan2016improved, xiao2017time, benesty2008microphone} and Generalized Eigenvalue (GEV) beamformer \cite{heymann2015blstm, grondin2020gev}. Although different types of beamformers \cite{benesty2008microphone, erdogan2016improved, heymann2015blstm, van2009speech} have different optimization and constraint conditions, most of their solutions are derived from the estimated speech and noise covariance matrices. These prior studies inspired us to use RNNs to directly learn the beamforming weights from the estimated speech and noise covariance matrices. 
%To overcome the above-mentioned issues,  
%of the mask based traditional beamformer, 
Hence, we recently proposed an all-deep-learning MVDR (ADL-MVDR) method \cite{zhang2020adl} which was superior to the traditional MVDR beamformer \cite{erdogan2016improved,xu2020neural}. In the ADL-MVDR \cite{zhang2020adl}, the matrix inversion and principal component analysis (PCA) operations of traditional MVDR are replaced by two RNNs with the estimated speech and noise covariance matrices as the input. 
%However, the ADL-MVDR is just one type of RNN-based beamformers. 
In this work, we propose more advanced and generalized RNN-based beamformers (RNN-BFs).
%However, more advanced and generalized RNN-BFs will be proposed in this work. %RNNs could model the temporal structure among frames so that it adaptively predicts the frame-wise beamforming weights. 
%However, it is not clear whether the advantage of the ADL-MVDR is coming from the estimated covariance matrices or following the MVDR solution formula.
%Significant residual noise reduction capability and better speech recognition accuracy were achieved \cite{zhang2020adl}. 
Note that there were also several other learning based beamforming methods \cite{xiao2016deep,meng2017deep}  which yielded worse performance than the traditional mask-based MVDR \cite{erdogan2016improved} approach due to the lack of explicitly using the speech and noise covariance matrices information \cite{xiao2016deep}. 

In this work, three contributions are made to further improve the ADL-MVDR \cite{zhang2020adl} beamformer. First, we propose a RNN-based GEV (RNN-GEV) beamformer, where it achieves slightly better performance than the ADL-MVDR \cite{zhang2020adl}. It indicates that the RNNs could also be incorporated into other traditional beamforming algorithms. Second, a generalized RNN beamformer (GRNN-BF) is proposed and it is superior to the RNN-GEV and the ADL-MVDR \cite{zhang2020adl}. The GRNN-BF directly learns the frame-level beamforming weights from covariance matrices without following conventional beamformers' solutions. It suggests that the GRNN-BF could learn a better beamforming solution by automatically accumulating the covariance matrices across history frames. 
%frame-level beamforming weights from the covariance matrices, without following conventional beamformers' solutions. 
%implement several other types of beamformers, including the RNN-based GEV beamformer and another generalized RNN beamformer (GRNN-BF). The GRNN-BF is demonstrated to achieve the best performance and have the capability to learn the beamforming weights directly from the 
%Most of traditional beamformers' solutions are based on the speech and noise covariance matrices \cite{benesty2008microphone, erdogan2016improved, heymann2015blstm, van2009speech}. RNNs could learn a better solution from the speech and noise covariance matrices directly. Second, RNNs could automatically accumulate the statistics of covariance matrices across history frames and predict the time-varying beamforming weights, considering that RNNs have the temporal modeling capability. Our proposed spatio-temporal RNN beamformer with time varying beamforming weights could be better than the time-invariant beamforming weights. 
Finally, the layer normalization \cite{ba2016layer} is proposed to replace the commonly used mask normalization \cite{boeddeker2018exploring, xiao2016deep, xu2020neural, zhang2021end} on the covariance matrices. The layer normalization is more flexible than the mask normalization and it can achieve better performance. These improvements make our proposed GRNN-BF perform the best in terms of PESQ, SNR and word error rate (WER) comparing to the traditional MVDR beamformers \cite{xu2020neural} and the ADL-MVDR beamformer \cite{zhang2020adl}.
%However, it is not clear whether the advantage of the ADL-MVDR is coming from the calculated covariance matrices or following the MVDR formula. More analysis on the ADL-MVDR module will be conducted in this work. The ADL-MVDR is one type of RNN-based beamforming methods. Here, we propose an improved RNN-based beamforming approach. 
%There are three contributions in this work. 
%First, we demonstrate that the estimated speech and noise covariance matrices are the key factors. This finding is important because it paves the road to design more optimal deep learning based beamformers. Second, three types of generalized RNN beamformers (GRNN-BFs) are proposed. Leveraging on the success of RNNs to solve the matrix inverse and eigenvalue decomposition problems \cite{zhang2005design,wang1993recurrent,liu2008recurrent,wang2016recurrent}, any kinds of traditional beamformers (e.g., MVDR \cite{erdogan2016improved}, GEV \cite{heymann2016neural}, Multichannel Wiener filtering \cite{van2009speech}, etc.) could be solved by the RNNs. However, only RNN-MVDR was investigated \cite{zhang2005design}. The proposed GRNN-BF achieves the best performance among all methods. Finally, we not only generalize the beamforming formula, but also generalize the RNN-BF model from two branches to one.

The rest of this paper is organized as follows. In section \ref{sec:mask_traditional_bf}, traditional mask-based beamformers are described. Section \ref{sec:rnnbf} presents the proposed generalized RNN-based beamformers (GRNN-BFs). The experimental setups and results are provided in Section \ref{sec:exp} and \ref{sec:res}, respectively. Section \ref{sec:conc} concludes this paper.
%\section{Related work}
%\label{sec:related}

\begin{figure*}[htb]
	\begin{minipage}[b]{1\linewidth}
		\centering
		\centerline{\includegraphics[width=1.0\textwidth]{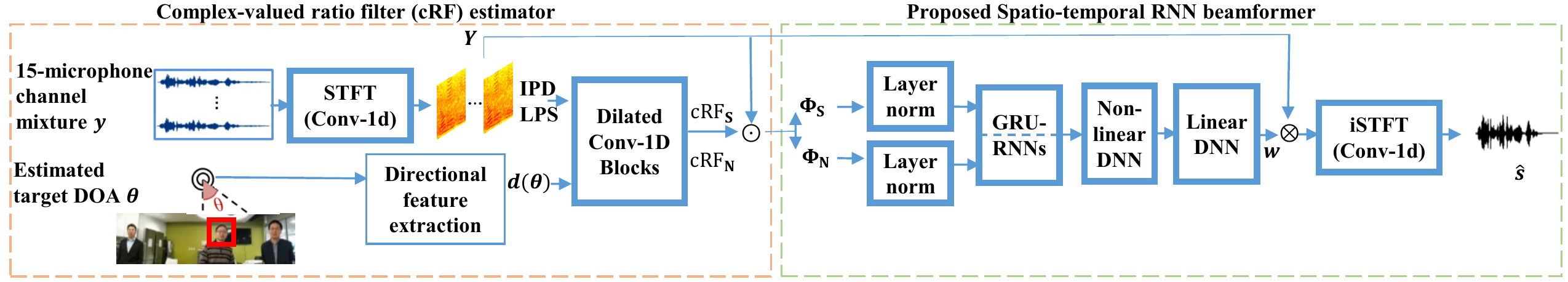}}
		%  \vspace{2.0cm}
		%\centerline{(a) Result 1}\medskip
	\end{minipage}
	\caption{The system framework includes the dilated Conv-1D blocks based complex-valued ratio filter (cRF) estimator and the proposed spatio-temporal RNN beamformer (RNN-BF). 
	The cRF estimator is actually a Conv-TasNet variant \cite{luo2019conv} with a fixed STFT encoder \cite{gu2020multi}). 
	$\odot$ indicates the complex-domain multiplication to estimate the multi-channel speech $\hat{\mathbf S}$ and noise $\hat{\mathbf N}$ through cRF (as shown in Eq. (\ref{eq:est_s})). $\otimes$ is the matrix multiplication of beamforming (see Eq. (\ref{eq:beamforming_filtering})). The speech covariance matrix ${\bf{{\Phi}}}_{\textbf{S}}$, noise covariance matrix ${\bf{{\Phi}}}_{\textbf{N}}$ and beamforming weights $\textbf{w}$ are complexed-valued variables, their real parts and imaginary parts are reshaped and concatenated together.  
	%$\odot$ and $\otimes$ indicate the operations expressed in Eq. (6) and (9), respectively.
	The time-domain scale-invariant SNR (Si-SNR) loss \cite{luo2019conv} is applied for end-to-end training.
    } %The front-end mask estimator and the back-end acoustic model are first separately trained. Then the joint training is performed by concatenating them together.}% with the beamforming module. }
	\label{fig:overview_system}
\end{figure*}

\section{Traditional mask-based beamformers}
\label{sec:mask_traditional_bf}
This section describes the traditional mask-based beamformers. 
%The proposed spatio-temporal RNN beamformers will be presented in Sec. \ref{sec:rnnbf}.
Given the $M$-channel speech mixture $\textbf{y}=[\textbf{y}_1, \textbf{y}_2,...,\textbf{y}_M]$, the corresponding $M$-channel target speaker's speech and the noise (the sum of interfering speakers' speech and background noise) waveforms are denoted as $\textbf{s}$ and $\textbf{n}$, respectively. After applying short-time Fourier transform (STFT), we have $\textbf{Y}, \textbf{S}, \textbf{N}$ in the time-frequency (T-F) domain,
\begin{equation}
\textbf{Y}(t,f)=\textbf{S}(t,f)+\textbf{N}(t,f)
\end{equation}
where $(t, f)$ indicates the time and frequency indices of the T-F domain variables. 
In conventional mask-based beamforming \cite{heymann2016neural,erdogan2016improved,xiao2017time,xu2020neural}, a neural network is used to predict the real-valued speech mask $\text{RM}_\textbf{S}$ and the real-valued noise mask $\text{RM}_\textbf{N}$.
%the target speech and noise covariance matrices are calculated through the NN-predicted masks. 
Then the speech covariance matrix ${\bf{{\Phi}}}_{\textbf{S}}$ is calculated with the predicted speech mask $\text{RM}_\textbf{S}$,
\begin{equation}
\label{eq:traditional_phi}
{\bf{{\Phi}}}_{\textbf{S}}(f)=\frac{\sum_{t=1}^{T}{\text{RM}}^2_\textbf{S}(t,f)\textbf{Y}(t,f)\textbf{Y}^{\sf H}(t,f)}{\sum_{t=1}^{T}{\text{RM}}^2_\textbf{S}(t,f)}
\end{equation}
Where $T$ stands for the total number of frames in a chunk. $\sf H$ is the Hermitian transpose. The noise covariance matrix  ${\bf{{\Phi}}}_{\textbf{N}}$ could be calculated in the same way with the noise mask $\text{RM}_\textbf{N}$. 
The MVDR solution \cite{erdogan2016improved} can be derived as,
\begin{equation} \small
\label{eq:MVDR_solution}
\mathbf{w}_\text{MVDR}(f) =\frac{\mathbf{\Phi}_{\mathbf{N}}^{-1}(f) \mathbf{\boldsymbol{v}}(f)}{\mathbf{\boldsymbol{v}^{\sf H}}(f) \mathbf{\Phi}_{\mathbf{N}}^{-1}(f) \mathbf{\boldsymbol{v}}(f)}, \quad \mathbf{w}_\text{MVDR}(f) \in \mathbb{C}^{M}
\end{equation}
where $\mathbf{\boldsymbol{v}}(f)$ represents the steering vector at $f$-th frequency bin. $\mathbf{\boldsymbol{v}}(f)$ could be derived by applying PCA on ${\bf{{\Phi}}}_{\textbf{S}}(f)$, namely $\mathbf{\boldsymbol{v}}(f)=\mathcal{P}\{{\bf{{\Phi}}}_{\textbf{S}}(f)\}$. %$\mathbf{w}_\text{MVDR}(f)$ is a $C$-channel complex-valued vector at $f$-th frequency bin. 
Another type of commonly used beamformer is generalized eigenvalue (GEV) \cite{heymann2016neural}, where its optimal solution is the generalized principle component \cite{heymann2016neural,grondin2020gev}, 
%\begin{equation}
%\label{eq:gev_formula}
%\mathbf{w}_\text{GEV}(f) %=\underset{\mathbf{w}}{\mathrm{argmax}}\frac{\mathbf{w}^{\sf H}(f) %{\bf{{\Phi}}}_{\textbf{S}}(f) \mathbf{w}(f)}{\mathbf{w}^{\sf H}(f) %{\bf{{\Phi}}}_{\textbf{N}}(f) \mathbf{w}(f)}
%\end{equation}
%This optimization problem leads to the generalized eigenvalue decomposition (GEVD) \cite{heymann2016neural}. The optimal solution is the generalized principal component \cite{heymann2016neural,grondin2020gev},
\begin{equation}
\label{eq:gev_solution}
\mathbf{w}_\text{GEV}(f) =\mathcal{P}\{ {\bf{{\Phi}}}^{-1}_{\textbf{N}}(f) {\bf{{\Phi}}}_{\textbf{S}}(f)\}, \quad \mathbf{w}_\text{GEV}(f) \in \mathbb{C}^{M}
\end{equation}
However, the beamforming weights $\mathbf{w}$ above are usually chunk-level \cite{heymann2016neural,erdogan2016improved,xiao2017time,xu2020neural}, which is not optimal for each frame. Furthermore, the matrix inversion involved in Eq. (\ref{eq:MVDR_solution}) and Eq. (\ref{eq:gev_solution}) has the numerical instability problem \cite{zhang2021end, chakrabarty2015numerical, lim2017numerical, zhao2012fast}. Note that we have already applied the diagonal loading technique \cite{mestre2003diagonal, chakrabarty2015numerical, zhang2021end} to alleviate this problem in our MVDR baselines.

On the other hand, although the MVDR and GEV are two different beamformers, their solutions are both derived from the speech and noise covariance matrices, namely ${\bf{{\Phi}}}_{\textbf{S}}$ and ${\bf{{\Phi}}}_{\textbf{N}}$. This is also our motivation to use RNNs to directly learn the beamforming weights from ${\bf{{\Phi}}}_{\textbf{S}}$ and ${\bf{{\Phi}}}_{\textbf{N}}$.
% Another defect is that most of these methods' solutions are in the chunk-level rather than the frame-wise. 

\section{Proposed generalized RNN beamformer} % (GRNN-BF)}
\label{sec:rnnbf}
We aim to extract the target speaker's speech from the multi-channel multi-talker overlapped mixture. As shown in Fig. \ref{fig:overview_system}, the whole system consists of a complex-valued ratio filter (cRF) estimator and the proposed spatio-temporal RNN beamformer. 
%The input to the cRF estimator is log-power spectra (LPS), interaural phase difference (IPD) \cite{tan2020audio} and the directional feature $d(\theta)$ \cite{chen2018multi} which is calculated based on the estimated target speaker's DOA $\theta$. The use of the target speaker's DOA will guide the model to extract the target speaker's speech. 
The predicted cRFs are used to calculate the covariance matrices. Then the proposed GRNN-BF learns the beamforming weights from the covariance matrices. The details of the input features and the cRF estimator will be described in Sec. \ref{sec:exp}. Our proposed GRNN-BF will be first illustrated here.

%\subsection{All deep learning MVDR (ADL-MVDR)}
%\subsection{Proposed spatio-temporal RNN beamformer}
Recently we proposed the ADL-MVDR method \cite{zhang2020adl} which is superior to the traditional mask-based MVDR beamformers \cite{erdogan2016improved, xiao2017time, xu2020neural}. The ADL-MVDR uses two RNNs to replace the matrix inversion and PCA in MVDR solution (defined in Eq. (\ref{eq:MVDR_solution})). Here we explore to use RNNs to implement the GEV beamformer (defined in Eq. (\ref{eq:gev_solution})) and another more generalized RNN beamformer. Layer normalization \cite{ba2016layer} is also proposed to replace the commonly used mask normalization \cite{boeddeker2018exploring, xiao2016deep, zhang2021end}, which is applied on the covariance matrices.

\subsection{Layer normalization on covariance matrix}
Before we use RNNs to learn the beamforming weights, the speech and noise covariance matrices should be first estimated. Real-valued masks \cite{boeddeker2018exploring}, complex-valued ratio mask (cRM) \cite{williamson2015complex, xu2020neural} or complex-valued ratio filter (cRF) \cite{mack2019deep,zhang2020adl} could be used to estimate the speech and noise. In our previous ADL-MVDR work \cite{zhang2020adl}, the cRF \cite{mack2019deep} is demonstrated to be better than the cRM \cite{williamson2015complex}. The cRF \cite{mack2019deep} is $K \times K$ size cRM \cite{williamson2015complex} by using nearby $K \times K$ T-F bins around $(t,f)$. 
%It is defined on the $(2K+1)$x$(2K+1)$ neighboring context (i.e., surrounding $(2K+1)$ frames and $(2K+1)$ frequency bins of T-F unit $(t,f)$). 
With the speech $\text{cRF}_\textbf{S}(t,f)$, the  multi-channel target speech is estimated as,
\begin{equation} \footnotesize
\label{eq:est_s}
\hat{\mathbf S}(t,f)=\sum_{\tau_1=-K}^{\tau_1=K} \sum_{\tau_2=-K}^{\tau_2=K} \text{cRF}_\textbf{S}(t+\tau_1,f+\tau_2) * \mathbf{Y}(t+\tau_1,f+\tau_2)
\end{equation}
%If $K=0$, then cRF \cite{mack2019deep} is exactly the same with cRM \cite{williamson2015complex}. 
%$\hat{\mathbf S}$ is the estimated multi-channel speech by multiplying the speech $\text{cRF}_\textbf{S}$ with $\textbf{Y}$ in the complex domain \cite{zhang2020adl}. 
Then the frame-wise speech covariance matrix is calculated as,
\begin{equation}\small
\label{eq:cov_matrix_ss}
{\bf{{\Phi}}}_{\textbf{S}}(t,f)=\frac{\hat{\mathbf S}(t,f) \hat{\mathbf S}^{\sf H}(t,f)} {\sum_{t=1}^{T} \text{cRM}_\textbf{S}^{\sf H}(t,f) \text{cRM}_\textbf{S}(t,f)}
\end{equation}
Where $\text{cRM}_\textbf{S}(t,f)$ is the center unit of the speech $\text{cRF}_\textbf{S}(t,f)$. Given the noise $\text{cRF}_{\textbf{N}}(t,f)$, the estimated multi-channel noise $\hat{\textbf{N}}(t,f)$ and the frame-wise noise covariance matrix ${\bf{{\Phi}}}_{\textbf{N}}(t,f)$ could be estimated in the same way. Different from Eq. (\ref{eq:traditional_phi}) where the covariance matrix are averaged over a chunk of frames in the traditional mask-based MVDR, the covariance matrices here are frame-wise. This is because the covariance matrices are later fed into RNNs where the unidirectional RNN could automatically accumulate the statistics of covariance matrices across history frames in a recursive way. Note that the denominator in Eq. (\ref{eq:cov_matrix_ss}) is the commonly used mask normalization \cite{boeddeker2018exploring, xiao2016deep, xu2020neural, zhang2021end} to normalize the covariance matrix. %This normalization is important for the training, especially with using the scale-invariant SNR (Si-SNR) loss \cite{luo2019conv}. 
In this work, we propose to use the layer normalization \cite{ba2016layer} to normalize the covariance matrices to achieve better performance.
\begin{equation}\small
\label{eq:ln_cov_matrix_ss}
{\bf{{\Phi}}}_{\textbf{S}}(t,f)=\text{LayerNorm}(\hat{\mathbf S}(t,f) \hat{\mathbf S}^{\sf H}(t,f))
\end{equation}
Where the layer normalization \cite{ba2016layer} applies per-element scale and bias with learnable affine transform, which is more flexible than the mask normalization. Another layer normalization is also  adopted for ${\bf{{\Phi}}}_{\textbf{N}}(t,f)$.

%%%
%\begin{equation}\small
%\label{eq:cov_matrix_ss}
%{\bf{{\Phi}}}_{\textbf{SS}}(t,f)={\hat{\mathbf S}(t,f) \hat{\mathbf S}^{\sf H}(t,f)}
%\end{equation}

%\begin{equation}\small
%\label{eq:cov_matrix_ss}
%\text{att}(t,f)=\text{RNN}({\text{cRF}^{\sf H}(t,f) \text{cRF}(t,f)})
%\end{equation}

%\begin{equation}\small
%\label{eq:cov_matrix_ss}
%{\bf{{\Phi}}}_{\textbf{SS}}(t,f)=\frac{\hat{\mathbf S}(t,f) \hat{\mathbf S}^{\sf H}(t,f)} {\text{att}(t,f)}
%\end{equation}
%%%

%The temporal sequence of speech and noise covariance matrices are fed into two RNNs to hypothetically learn the steering vector and the matrix inversion, respectively.
%\begin{align} \small
%\hat{\mathbf{\boldsymbol{v}}} (t,f) = \text{RNN} ( {\bf{{\Phi}}}_{\textbf{S}}(t,f)) \\
%\small {\bf{{\hat \Phi}}}_{\textbf{N}}^{-1}(t,f) = \text{RNN} ( {\bf{{{\Phi}}}}_{\textbf{N}}(t,f))
%\end{align}
%The uni-directional RNNs could automatically accumulate and update the covariance matrices from the history frames. Finally, the frame-wise beamforming weights are,
%\setlength{\abovedisplayskip}{7pt}
%\setlength{\belowdisplayskip}{7pt}
%\begin{equation}\small
%\label{eq:adlMVDR_solution}
%\mathbf{w}_\text{ADL-MVDR}(t,f) =\frac{\mathbf{\hat \Phi}_{\mathbf{N}}^{-1}(t,f) \boldsymbol{\hat v}(t,f)}{\mathbf{\boldsymbol{\hat v}^{\sf H}}(t,f) \mathbf{\hat \Phi}_{\mathbf{N}}^{-1}(t,f) \mathbf{\boldsymbol{\hat v}}(t,f)}
%\end{equation}
%\setlength{\abovedisplayskip}{0.1pt}
%\setlength{\belowdisplayskip}{1pt}

\subsection{Spatio-temporal RNN GEV beamformer}
\label{sec:RNN-GEV}
%However, it is not clear whether the key factor in ADL-MVDR is the calculated covariance matrices or following the MVDR formula (as shown in Eq. (\ref{eq:adlMVDR_solution})). Here we propose different types of generalized RNN beamformers (GRNN-BFs).
%, including GEV beamformer and a direct weight prediction beamformer. A generalized RNN beamformer is also proposed. 
Similar to the ADL-MVDR \cite{zhang2020adl}, here the proposed spatio-temporal RNN GEV beamformer (RNN-GEV) also takes the estimated target speech covariance matrix ${\bf{{\Phi}}}_{\textbf{S}}(t,f)$ and the noise
%(including the interfering speech and additive noise)
covariance matrix ${\bf{{\Phi}}}_{\textbf{N}}(t,f)$ as the input to predict the frame-wise beamforming weights. 
%\subsubsection{RNN-based GEV beamformer (RNN-GEV)}
Following the solution of the traditional GEV beamformer defined in Eq. (\ref{eq:gev_solution}), we reformulate its form in the RNN-based beamforming framework as,
\begin{align} \small
{\bf{\hat \Phi}}_{\textbf{N}}^{-1}(t,f) &= \text{RNN} ( {\bf{{\Phi}}}_{\textbf{N}}(t,f)) \label{eq:rnn_matrix_inverse} \\
\small {\bf{{\hat \Phi}}}_{\textbf{S}}(t,f) &= \text{RNN} ( {\bf{{\Phi}}}_{\textbf{S}}(t,f)) \label{eq:rnn_matrix_psds} \\
\small \mathbf{w}_\text{RNN-GEV}(t,f) &= \text{DNN}( {{\bf{\hat \Phi}}_{\textbf{N}}^{-1}(t,f) \bf{{\hat \Phi}}}_{\textbf{S}}(t,f)) \\
\small \hat{\mathbf{S}}(t,f)&=(\mathbf{w}_\text{RNN-GEV}(t,f))^{\sf H}\mathbf{Y}(t,f)
\end{align}
where $\mathbf{w}_\text{RNN-GEV}(t,f) \in \mathbb{C}^{M}$. ${\bf{{\hat \Phi}}}_{\textbf{S}}(t,f)$ is the accumulated speech covariance matrix from the history frames by leveraging on the temporal modeling capability of RNNs. ${\bf{\hat \Phi}}_{\textbf{N}}^{-1}(t,f)$ is assumed to be the matrix inversion of ${\bf{ \Phi}}_{\textbf{N}}(t,f)$. Instead of using the actual generalized PCA (as in Eq. (\ref{eq:gev_solution})), a deep neural network (DNN) is utilized to calculate the beamforming weights for RNN-GEV. Hinton et al \cite{hinton2006reducing} shows that the DNN has the ability to conduct the non-linear generalized PCA. %Note that no linear algebra decompositions are used here.

%\subsubsection{Generalized RNN beamformer I (GRNN-BF-I)}
%Similar to GRNN-GEV, we also propose another generalized RNN beamformer (GRNN-BF-I). As shown in Fig. \ref{fig:overview_system}, ${\bf{\hat \Phi}_{\textbf{N}}^{-1}}(t,f)$ and ${\bf{{\hat \Phi}}}_{\textbf{S}}(t,f)$ are concatenated rather then multiplied as the GRNN-GEV did. 
%\begin{align} \small
%%{\bf{\hat \Phi}}_{\textbf{NN}}^{-1}(t,f) = \text{RNN} ( {\bf{{\Phi}}}_{\textbf{NN}}(t,f)) \\
%%{\bf{{\hat \Phi}}}_{\textbf{SS}}(t,f) = \text{RNN} ( {\bf{{\Phi}}}_{\textbf{SS}}(t,f)) \\
%\mathbf{w}_\text{GRNN-BF-I}(t,f) =\text{DNN}([{{\bf{\hat \Phi}}_{\textbf{N}}^{-1}(t,f), \bf{{\hat \Phi}}}_{\textbf{S}}(t,f)])
%\end{align}
%where $\mathbf{w}_\text{GRNN-BF-I}(t,f) \in \mathbb{C}^{C}$. 
%%To further get rid of the traditional beamforming forms, here 
%We try to predict the beamforming weights directly, without following any traditional beamformers' formulas (e.g., MVDR or GEV). 
%%Note that the physical meaning of ${\bf{\hat \Phi}}_{\textbf{N}}^{-1}(t,f)$ and ${\bf{{\hat \Phi}}}_{\textbf{S}}(t,f)$ (after processed by RNNs as in Eq. (\ref{eq:rnn_matrix_inverse}) and (\ref{eq:rnn_matrix_psds})) might be changed in GRNN-BF-I due to the more flexible optimization. 
%All of the covariance matrices and beamforming weights are complex-valued, and we concatenate the real and imaginary parts of any complex-valued matrices or vectors in the whole work.

\subsection{Generalized spatio-temporal RNN beamformer}
Finally, we propose a more generalized spatio-temporal RNN beamformer (GRNN-BF) without following any traditional beamformers' solutions. This is motivated by that, different beamformers (e.g., MVDR and GEV) have different solutions but almost all solutions are derived from the speech and noise covariance matrices. The neural networks could be able to learn a better solution from the speech and noise covariance matrices.
%As shown in Fig. \ref{fig:overview_system}, the model structure for predicting the beamforming weights from the calculated covariance matrices ${\bf{{\Phi}}}_{\textbf{S}}(t,f)$ and ${\bf{{\Phi}}}_{\textbf{N}}(t,f)$ are further generalized from two branches to one branch.
The RNN-GEV and the ADL-MVDR \cite{zhang2020adl} both have two RNNs to deal with the target speech covariance matrix $\bf{\hat \Phi}_{\textbf{S}}(t,f)$ and the noise covariance matrix ${\bf{{\hat \Phi}}}_{\textbf{N}}(t,f)$, respectively. But GRNN-BF here uses only one unified RNN-DNN model to predict the frame-level beamforming weights directly.
\begin{align} \small
\mathbf{w}_\text{GRNN-BF}(t,f) &= \text{RNN-DNN}([{{\bf{ \Phi}}_{\textbf{N}}(t,f), \bf{{ \Phi}}}_{\textbf{S}}(t,f)]) \\
\label{eq:beamforming_filtering}
\hat{\mathbf{S}}(t,f)&=(\mathbf{w}_\text{GRNN-BF}(t,f))^{\sf H}\mathbf{Y}(t,f)
\end{align}
where $\mathbf{w}_\text{GRNN-BF}(t,f) \in \mathbb{C}^{M}$. The input for the RNN-DNN is the concatenated tensor of ${\bf{ \Phi}_{\textbf{N}}}(t,f)$ and ${\bf{ \Phi}_{\textbf{S}}}(t,f)$. 
%$\text{Real}({\bf{ \Phi}}_{\textbf{S}}(t,f))$, $\text{Imag}({\bf{ \Phi}}_{\textbf{S}}(t,f))$, $\text{Real}({\bf{ \Phi}}_{\textbf{N}}(t,f))$ and $\text{Imag}({\bf{ \Phi}}_{\textbf{N}}(t,f))$. 
All of the covariance matrices and beamforming weights are complex-valued, and we concatenate the real and imaginary parts of any complex-valued tensors in the whole work.

\section{Dataset and experimental setup}
\label{sec:exp}

\begin{table*}[htbp] \small
	\centering
	\caption{PESQ, Si-SNR(dB) \cite{luo2019conv}, SDR(dB) and WER($\%$) results among Conv-TasNet with STFT \cite{gu2020multi}, several MVDRs and proposed GRNN-BF systems. "MN" and "LN" denote mask normalization and layer normalization on the covariance matrices, respectively.}
\begin{tabular}{l|c|c|c|c|c|c|c|c|c|c|c}
	\hline 
	systems/metrics & \multicolumn{8}{c}{PESQ [-0.5, 4.5]}                          & \multicolumn{1}{|l|}{Si-SNR} & \multicolumn{1}{|l|}{SDR} & \multicolumn{1}{|l}{WER} \\
	\hline 
	& \multicolumn{4}{|c|}{Angle between target \& others} & \multicolumn{4}{|c|}{\# of overlapped speakers} &       &       &  \\
	\hline 
	& 0-15  & 15-45 & 45-90 & 90-180 & 1SPK  & 2SPK  & 3SPK  & Avg.  & Avg.  & Avg.  & Avg. \\
	\hline 
	Reverberant clean reference & 4.50  & 4.50  & 4.50  & 4.50  & 4.50  & 4.50  & 4.50  & 4.50  & $\infty$ & $\infty$ & 8.26 \\
	Mixture (overlapped speech+noise) & 1.88  & 1.88  & 1.98  & 2.03  & 3.55  & 2.02  & 1.77  & 2.16  & 3.39  & 3.50  & 55.14 \\
	\hline 
	Conv-TasNet with STFT (i) \cite{gu2020multi} & 2.75  & 2.95  & 3.12  & 3.09  & 3.98  & 3.06  & 2.76  & 3.10  & 12.50 & 13.01 & 22.07 \\
	MVDR w/ MN (ii) \cite{xu2020neural} & 2.55  & 2.77  & 2.96  & 2.89  & 3.82  & 2.90  & 2.55  & 2.92  & 11.31 & 12.58 & 15.91 \\
	Multi-tap MVDR w/ MN (iii) \cite{xu2020neural} & 2.67  & 2.95  & 3.15  & 3.10  & 3.92  & 3.06  & 2.72  & 3.08  & 12.66 & 14.04 & 13.52 \\
	ADL-MVDR w/ MN (iv) \cite{zhang2020adl} & 3.04  & 3.30  & 3.48  & 3.48  & 4.17  & 3.41  & 3.07  & 3.42  & 14.80 & 15.45 & 12.73 \\
	\hline 
	\textbf{Prop. RNN-GEV w/ MN} (v) & 3.11  & 3.36  & 3.55  & 3.54  & 4.19  & 3.48  & 3.14  & 3.48  & 15.34 & 15.88 & 12.07 \\
	\textbf{Prop. RNN-GEV w/ LN} (vi) & 3.15  & 3.39  & 3.57  & 3.56  & 4.19  & 3.51  & 3.17  & 3.51  & 15.55 & 16.07 & 11.75 \\
	\textbf{Prop. GRNN-BF w/ MN} (vii) & 3.17  & 3.40  & 3.58  & 3.59  & 4.21  & 3.53  & 3.19  & 3.52  & 15.48 & 16.03 & 11.86 \\
	%\textbf{Prop. Chunk GRNN-BF w/ LN} (viii) & 2.61  & 2.86  & 3.06  & 3.00  & 3.87  & 2.98  & 2.65  & 3.01  & 11.99 & 13.31 & 14.72 \\
	\textbf{Prop. GRNN-BF w/ LN} (viii) & \textbf{3.23} & \textbf{3.45} & \textbf{3.62} & \textbf{3.60} & \textbf{4.23} & \textbf{3.57} & \textbf{3.24} & \textbf{3.56} & \textbf{15.84} & \textbf{16.38} & \textbf{11.36} \\
	\hline 
\end{tabular}%

	\label{tab:pesq}%
\end{table*}%
%\subsection{Experimental setup}
\textbf{Dataset}: The methods are evaluated on the mandarin audio-visual corpus \cite{tan2020audio,gu2020multi}, which is collected from YouTube \cite{zhang2020multi}. The dataset has 205500 clean speech segments (about 200 hours) over 1500 speakers. The audio sampling rate is 16 kHz. 512-point of STFT is used to extract features along 32ms Hann window with 50\% overlap.
%% simulation
%The multi-talker multi-channel dataset are simulated in the similar way with our previous work \cite{tan2020audio,gu2020multi}. 
There are one to three overlapped speaking speakers in the simulated 15-channel mixture signal. The signal-to-interference ratio (SIR) is ranging from -6 to 6 dB. Noise with 18-30 dB SNR is added to all the 15-channel mixtures \cite{tan2020audio}. We use a 15-element non-uniform linear array. Based on the image-source simulation method \cite{habets2006room}, the simulated dataset contains 190000, 15000 and 500 multi-channel mixtures for training, validation and testing. % The speakers in the test set are unseen in the training set.
%The multi-channel signals are generated by convolving speech with RIRs simulated by image-source method \cite{habets2006room}.
The virtual acoustic room size is ranging from 4m-4m-2.5m to 10m-8m-6m.  %(length-width-height). 
The reverberation time T60 is sampled in a range of 0.05s to 0.7s. 
%PESQ, Si-SNR and SDR are used to evaluate our systems. 
%\setlength{\parskip}{0em}

\textbf{cRF estimator:} \label{sec:crf_est} we use the complex-valued ratio filter (cRF) \cite{zhang2020adl,mack2019deep} to estimate the covariance matrices. As shown in Fig. \ref{fig:overview_system}, the input to the cRF estimator includes a 15-channel mixture audio and a target direction of arrival (DOA) ($\theta$). From the multi-channel audio, log-power spectra (LPS) and interaural phase difference (IPD) \cite{tan2020audio} features are extracted. For the simulated data, the ground-truth target DOA is known. For the real-world scenario, we have the hardware where the $180^{\circ}$ wide-angle camera and the 15-linear microphone array are aligned \cite{xu2020neural}. Hence the target DOA ($\theta$) could be roughly estimated from the camera view by locating the target speaker's face (see our actual hardware demo website: \href{https://yongxuustc.github.io/grnnbf}{https://yongxuustc.github.io/grnnbf}). Then the DOA guided directional feature (DF) \cite{chen2018multi}, namely $d(\theta)$, is estimated by calculating the cosine similarity between the target steering vector $\mathbf{\boldsymbol{v}}$ and IPDs \cite{chen2018multi,gu2020multi}. Target DOA and $d(\theta)$ are speaker-dependent features which can be used to extract the target speech. LPS, IPD and $d(\theta)$ are merged and fed into a Conv-TasNet variant \cite{luo2019conv} with a fixed STFT encoder \cite{tan2020audio,gu2020multi}. A stack of eight successive dilated Conv-1D layers with 256 channels forms a network block and four blocks are piled together. The estimated cRF \cite{mack2019deep} size ($K \times K$) is empirically set to 3x3 \cite{mack2019deep,zhang2020adl}.
%, namely $K$ in Eq. (\ref{eq:est_s}) is set to one.

As for the RNN-BF module, the RNNs have 2-layer gated recurrent units (GRUs) with 500 hidden nodes. The non-linear DNN layer has 500 PReLU units. There are 30 linear units at the output DNN layer to predict the frame-wise beamforming weights. The model is trained in a chunk-wise mode with 4-second chunk size, using Adam optimizer. Initial learning rate is set to 1e-4. The objective is to maximize the time-domain scale-invariant source-to-noise ratio (Si-SNR) \cite{luo2019conv}. Pytorch 1.1.0 was used. Gradient norm is clipped with max norm 10. We evaluate the systems by using different metrics, including PESQ, Si-SNR (dB), signal-to-distortion ratio (SDR) (dB). A commercial general-purpose mandarin speech recognition Tencent API \cite{tencent_api} is used to test the ASR performance in WER. Note this work only focuses on speech separation and denoising without dereverberation. Hence the reverberant clean (without dereverberation) is used as the reference signal.

\section{Results and discussions}\label{sec:res}
%\subsubsection{Evaluations of the proposed GRNN-BF}
We evaluate the target speech separation performance in the overlapped multi-talker scenario. The spatial angle between the target speaker and others (interfering speakers) lies within 0-180$^\circ$. The more overlapped speakers and the smaller spatial angle would lead to more challenging separation tasks. The detailed PESQ scores across different scenarios (i.e., angle between the target speaker and other speakers; number of overlapped speakers) are presented in Table \ref{tab:pesq}. Other metrics are given with the average results.

\begin{figure}[htb]
	\begin{minipage}[b]{1.0\linewidth}
		\centering
		\centerline{\includegraphics[width=1.0\textwidth]{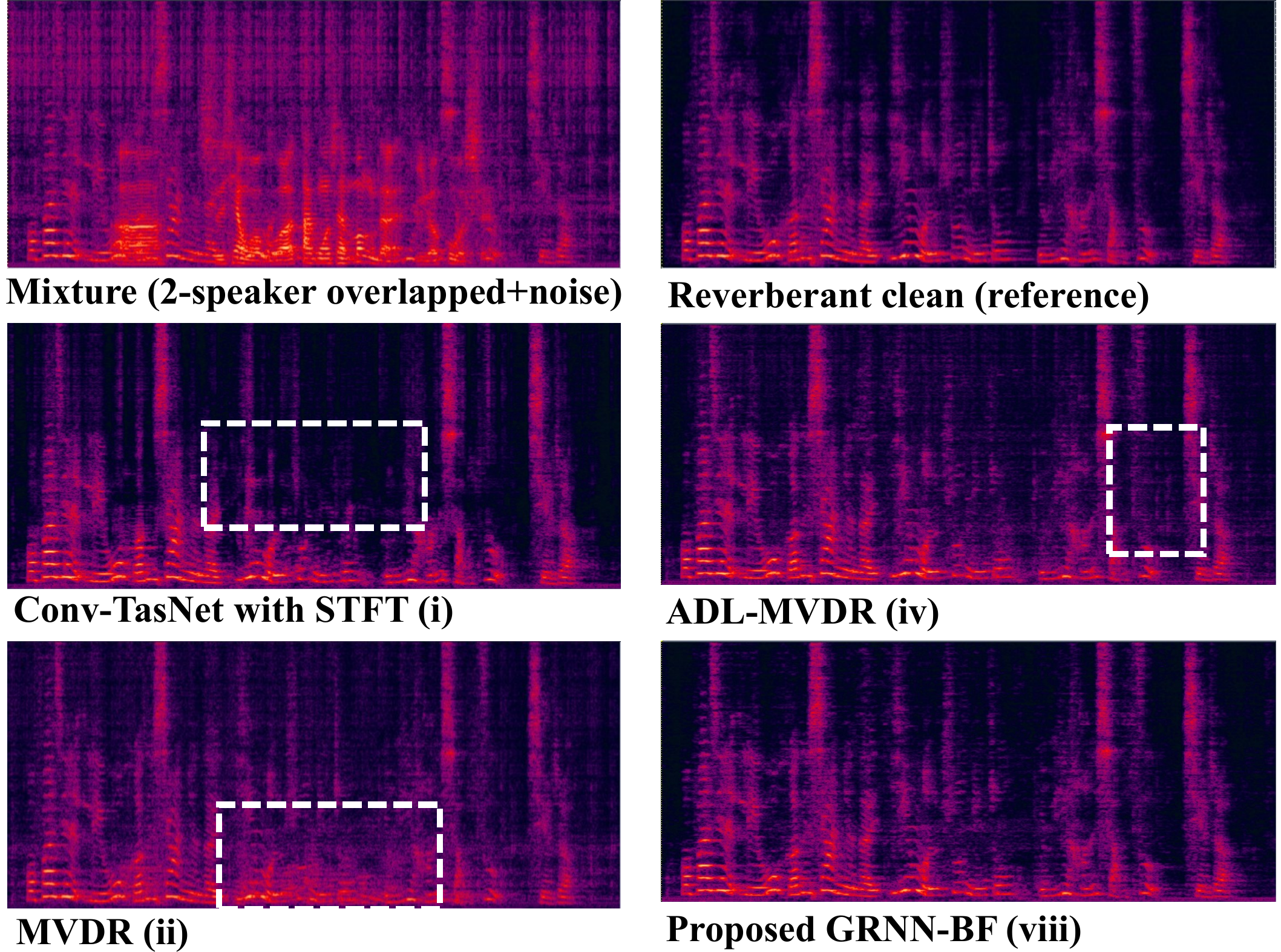}}
		%  \vspace{2.0cm}
		%\centerline{(a) Result 1}\medskip
	\end{minipage}
	\caption{Sample separated spectrograms of different target speech separation systems. More testing demos (including real-world recording demos to verify the generalization capability) could be found at: \href{https://yongxuustc.github.io/grnnbf}{https://yongxuustc.github.io/grnnbf}.}% The target DOA is 110$^\circ$.} %and the interfering speaker's DOA is 7$^\circ$. There is also some non-stationary additive noise.}
	\label{fig:spec}
\end{figure}

\textbf{GRNN-BF vs. traditional MVDRs}: Two traditional MVDR systems, MVDR with mask normalization (ii) \cite{xu2020neural,zhang2020adl} and multi-tap (i.e., [$t-1, t$]) MVDR with mask normalization (iii) \cite{xu2020neural,zhang2020adl} are compared here.  They also use the cRF estimator to calculate covariance matrices but replace the RNN-BF module (as shown in Fig. \ref{fig:overview_system}) with conventional MVDR or multi-tap MVDR \cite{xu2020neural} solutions. They both work reasonably well, e.g., the multi-tap MVDR (iii) achieves 13.52\% WER. However, the proposed GRNN-BF with mask normalization (vii) could obtain significantly better performance. The proposed GRNN-BF (vii) increases the average PESQ to 3.52 from 3.08 of multi-tap MVDR (iii) and 2.92 of MVDR (ii). The WER of the proposed GRNN-BF (vii) is better than the multi-tap MVDR (iii), i.e., 11.86 vs. 13.52. The corresponding Si-SNR and SDR are increased to 15.48 dB and 16.03 dB, respectively. Fig. \ref{fig:spec} also shows that the proposed GRNN-BF could estimate the spectra with less residual noise than traditional MVDR. The traditional MVDR has limited noise reduction capability \cite{habets2013two, xu2020neural}.  Finally, the proposed GRNN-BF with layer normalization (viii) achieves the best performance among all systems across all metrics. Moreover, the PESQ scores of our proposed GRNN-BF (viii) are above 3.2 at all scenarios, especially the two most difficult cases, namely small angle ($<$15$^\circ$) and three overlapped speakers (3SPK).
%The beam pattern in Fig. \ref{fig:beam} explains that the proposed GRNN-BF has a better beam focus on the target direction while suppressing most of the interfering noise from other directions. 
%Although the MVDR baselines (ii and iii) are worse than our proposed GRNN-BF system (viii), they still work well, e.g.,  As we have already used some techniques, e.g., the diagonal loading  \cite{mestre2003diagonal, chakrabarty2015numerical, zhang2021end}, to alleviate the numerical instability problem in traditional MVDRs.
%

%\begin{figure}[htb]
%	\begin{minipage}[b]{1.0\linewidth}
%		\centering
%		\centerline{\includegraphics[width=1.0\textwidth]{beam.pdf}}
%		%  \vspace{2.0cm}
%		%\centerline{(a) Result 1}\medskip
%	\end{minipage}
%	\caption{Beam patterns of MVDR and proposed GRNN-BF. The microphone array is 15-element non-uniform linear array.}% The target DOA is 110$^\circ$.} %and the interfering speaker's DOA is 7$^\circ$. There is also some non-stationary additive noise.}
%	\label{fig:beam}
%\end{figure}

\textbf{GRNN-BF vs. RNN-MVDR/GEV}: Our proposed RNN-GEV uses RNNs to implement the GEV beamformer following Eq. (\ref{eq:gev_solution}) while the ADL-MVDR \cite{zhang2020adl} following Eq. (\ref{eq:MVDR_solution}). With a more flexible structure (as shown in Sec. \ref{sec:RNN-GEV}), the proposed RNN-GEV (v) is slightly better than the ADL-MVDR (iv), e.g., PESQ: 3.48 vs. 3.42; WER: 12.07 vs. 12.73. However, the proposed GRNN-BF (vii) is better than both of them. Compared to the ADL-MVDR (iv), the proposed GRNN-BF (vii) further improves the average PESQ from 3.42 to 3.52 and the average Si-SNR from 14.80 dB to 15.48 dB. Fig. \ref{fig:spec} also shows that the proposed GRNN-BF can enhance the spectrogram with less residual noise than the ADL-MVDR. These results suggest that there is no need to follow any beamformers' solutions. The RNNs could learn a better solution from the speech and noise covariance matrices directly. The layer normalization is better than the mask normalization to normalize covariance matrices for both of the proposed RNN-GEV (vi) and GRNN-BF (viii).

\textbf{GRNN-BF vs. Conv-TasNet}: Conv-TasNet with a fixed STFT encoder \cite{gu2020multi} is our cRF estimator (as shown in Fig. \ref{fig:overview_system}), which is a variant of the original Conv-TasNet \cite{luo2019conv}. It is a purely ``black-box" neural network system with the same multi-channel input. It predicts the target speech as Eq. (\ref{eq:est_s}) defined. 
%by multiplying the predicted $\text{cRF}_\textbf{S}$ with the 0-th channel mixture $\textbf{Y}^{(0)}$. 
The proposed GRNN-BF with layer normalization (viii) beats the Conv-TasNet with STFT (i) by a large margin, i.e, PESQ: 3.56 vs. 3.10; Si-SNR: 15.84 vs. 12.50; WER: 11.36 vs. 22.07. The Conv-TasNet results in the worst WER 22.07\% among all systems due to the non-linear distortion which is quite common in most of purely neural network based speech separation systems \cite{du2014robust,luo2020distortion,xu2020neural}. This non-linear distortion can also be found in the separated spectrogram of Conv-TasNet in Fig. \ref{fig:spec}.

\textbf{Layer normalization vs. mask normalization}: As the denominator defined in Eq. (\ref{eq:cov_matrix_ss}), the mask normalization \cite{boeddeker2018exploring, xiao2016deep, xu2020neural, zhang2021end} on the covariance matrix are always applied to stabilize the training. However, the proposed layer normalization on the covariance matrix (as defined in Eq. (\ref{eq:ln_cov_matrix_ss})) is more flexible than the mask normalization. The proposed GRNN-BF with layer normalization (viii) obtains better performance than the GRNN-BF with mask normalization (vii), e.g., WER: 11.36 vs. 11.86; PESQ: 3.52 vs. 3.56.

%Fig. \ref{fig:spec} shows the example spectrograms and beam patterns of the MVDR and the proposed GRNN-BF-II. More demos (including real-world recording demos) could be found at: \href{https://yongxuustc.github.io/grnnbf}{https://yongxuustc.github.io/grnnbf}. The traditional cRF/mask based MVDR method has more residual noise (shown in the dashed rectangle). Its corresponding beam pattern is not sharp enough. However the GRNN-BF-II could significantly reduce most of the residual noise and keep the separated speech nearly distortionless (also demonstrated by the lowest WER 11.86\% in Table \ref{tab:pesq}.). %This is benefited from the adaptively predicted frame-wise beamforming weights of the proposed GRNN-BF-II. 
%The sharp beam pattern (selected from one of frames) of the proposed GRNN-BF-II in Fig. \ref{fig:spec_beam} also reflects its noise reduction capability.

\section{Conclusions}
\label{sec:conc}
In summary, we proposed a generalized RNN beamformer (GRNN-BF) that learns the frame-level beamforming weights directly from the estimated speech and noise covariance matrices. The layer normalization achieves better performance than the mask normalization for normalizing the speech and noise covariance matrices.
%The GRNN-BFs are beyond our previous ADL-MVDR work in the beamforming solution and model structure designs. We conclude the finding that the calculated speech and noise covariance matrices are the most important factors. There is no need to follow the formulas of any traditional beamformers. The GRNN-BF could automatically accumulate and update the covariance matrices from the history frames and predict the frame-wise beamforming weights directly. 
The proposed GRNN-BF with layer normalization achieves the best objective scores (PESQ, Si-SNR, SDR) and the lowest WER among all evaluated systems. It achieves relative 10.8\% and 16\% WER reduction against the prior art methods, ADL-MVDR \cite{zhang2020adl} and the conventional multi-tap MVDR \cite{xu2020neural}, respectively. Although we only tested the proposed GRNN-BF on the DOA-guided target speech separation task, it could also be used for general scenarios without DOA information, e.g., the multi-channel speech enhancement task or the permutation invariant training (PIT) \cite{yu2017permutation} based multi-channel speech separation task.
%In the future work, we will investigate the GRNN-BF's capability on the joint separation and dereverberation.

\pagebreak

\bibliographystyle{IEEEtran}

\bibliography{mybib}

% \begin{thebibliography}{9}
% \bibitem[1]{Davis80-COP}
%   S.\ B.\ Davis and P.\ Mermelstein,
%   ``Comparison of parametric representation for monosyllabic word recognition in continuously spoken sentences,''
%   \textit{IEEE Transactions on Acoustics, Speech and Signal Processing}, vol.~28, no.~4, pp.~357--366, 1980.
% \bibitem[2]{Rabiner89-ATO}
%   L.\ R.\ Rabiner,
%   ``A tutorial on hidden Markov models and selected applications in speech recognition,''
%   \textit{Proceedings of the IEEE}, vol.~77, no.~2, pp.~257-286, 1989.
% \bibitem[3]{Hastie09-TEO}
%   T.\ Hastie, R.\ Tibshirani, and J.\ Friedman,
%   \textit{The Elements of Statistical Learning -- Data Mining, Inference, and Prediction}.
%   New York: Springer, 2009.
% \bibitem[4]{YourName17-XXX}
%   F.\ Lastname1, F.\ Lastname2, and F.\ Lastname3,
%   ``Title of your INTERSPEECH 2021 publication,''
%   in \textit{Interspeech 2021 -- 20\textsuperscript{th} Annual Conference of the International Speech Communication Association, September 15-19, Graz, Austria, Proceedings, Proceedings}, 2020, pp.~100--104.
% \end{thebibliography}

\end{document}